\pdfoutput=1
\documentclass{statsoc}
\usepackage{hyperref}
\usepackage{bm,graphicx,color,amsbsy,amsmath,amsfonts,amssymb,
                   subcaption}
                
\usepackage[left=7mm,right=15mm]{geometry} 
\usepackage{vmargin} 

\usepackage{amsmath,amssymb,hyperref}
\usepackage{natbib}
\usepackage{pdfsync}
\usepackage{float}

\newcommand{\SSs}{\ensuremath{\mbox{\scriptsize \sf SS } }}
\newcommand{\GS}{\ensuremath{\mbox{\scriptsize \sf GS } }}
\newcommand{\BrS}{\ensuremath{\mbox{\scriptsize \sf BrS } }}

\title[Partially-latent class models for etiology estimation]{Partially-Latent Class Models (pLCM) for Case-Control\\ Studies of Childhood Pneumonia Etiology}
\author{Zhenke Wu}
\address{Department of Biostatistics, Johns Hopkins University, Baltimore, MD 21205, USA\\
Email: zhwu@jhu.edu}
\email{}
\author{Maria Deloria-Knoll}
\address{Department of International Health, Johns Hopkins University, Baltimore, MD 21205, USA}
\author{Laura L. Hammitt}
\address{Department of International Health, Johns Hopkins University, Baltimore, MD 21205, USA}
\author[Z.Wu {\it et al.}]{Scott L. Zeger}
\address{Department of Biostatistics, Johns Hopkins University, Baltimore, MD 21205, USA}
\author[Z.Wu {\it et al.}]{for the PERCH Core Team}

\begin{document}
\maketitle
\begin{abstract}
{In population studies on the etiology of disease, one goal is the estimation of the fraction of cases attributable to each of several causes. For example, pneumonia is a clinical diagnosis of lung infection that may be caused by viral, bacterial, fungal, or other pathogens. The study of pneumonia etiology is challenging because directly sampling from the lung to identify the etiologic pathogen is not standard clinical practice in most settings. Instead, measurements from multiple peripheral specimens are made. This paper introduces the statistical methodology designed for estimating the \textit{population etiology distribution} and the \textit{individual etiology probabilities} in the Pneumonia Etiology Research for Child Health (PERCH) study of $9,500$ children for $7$ sites around the world. We formulate the scientific problem in statistical terms as estimating the mixing weights and latent class indicators under a partially-latent class model (pLCM) that combines heterogeneous measurements with different error rates obtained from a case-control study. We introduce the pLCM as an extension of the latent class model. We also introduce graphical displays of the population data and inferred latent-class frequencies. The methods are tested with simulated data, and then applied to PERCH data. The paper closes with a brief description of extensions of the pLCM to the regression setting and to the case where conditional independence among the measures is relaxed.}
\end{abstract}
\vspace{-1cm}

\section{Introduction}
Identifying the pathogens responsible for infectious diseases in a population poses significant statistical challenges. 
Consider the measurement problem in the Pneumonia Etiology Research for Child Health (PERCH), a case-control study that has enrolled $9,500$ children from 7 sites around the world. Pneumonia is a clinical syndrome that develops because of an infection of the lung tissue by bacteria, viruses, mycobacteria or fungi \citep{Levine2012}. The appropriate treatment and public health control measures vary by pathogen. Which pathogen is infecting the lung usually cannot be directly observed and must therefore be inferred from multiple peripheral measurements with differing error rates. The primary goals of the PERCH study are to integrate the multiple sources of data to: (1) attribute a particular case's lung infection to a pathogen, and (2) estimate the prevalences of the etiologic pathogens in a population of children that met clinical pneumonia definitions.

\begin{figure}[!p]
\begin{center}
\includegraphics[width=\textwidth]{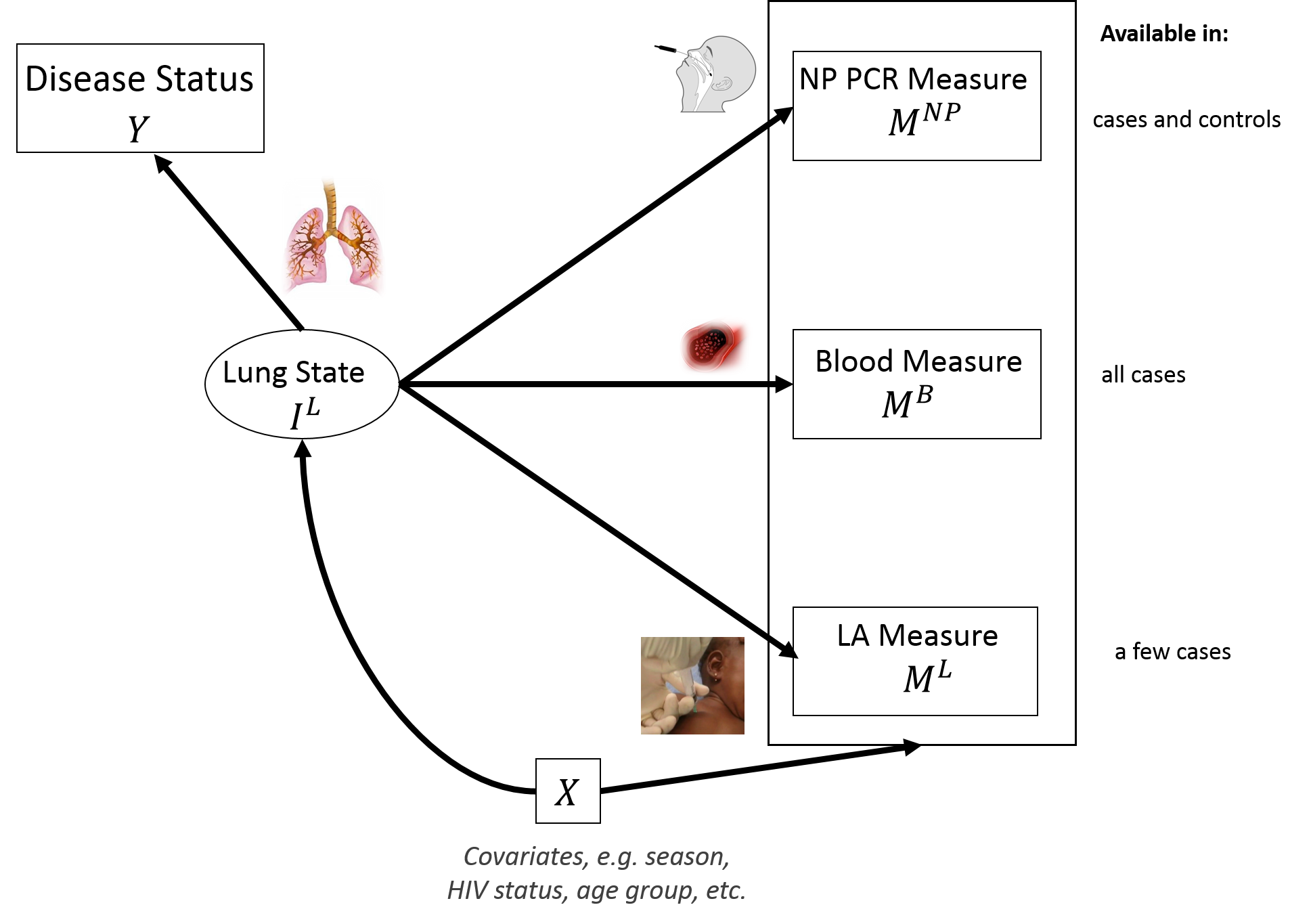}
\end{center}
\vspace{-0.1in}
\caption[Directed acyclic graph (DAG) illustrating relationships among lung infection state ($I^L$), imperfect lab measurements on the presence/absence of each of  a list of pathogens at each site($M^{NP}$, $M^{B}$ and $M^{L}$), disease outcome, and covariates ($X$).]{Directed acyclic graph (DAG) illustrating relationships among lung infection state ($I^L$), imperfect lab measurements on the presence/absence of each of a list of pathogens at each site($M^{NP}$, $M^{B}$ and $M^{L}$), disease outcome ($Y$), and covariates ($X$).}
\label{fig:basicstructure}
\end{figure}

The basic statistical framework of the problem is pictured in Figure \ref{fig:basicstructure}. The disease status is determined by clinical examination including chest X-ray \citep{Deloria2012}. The known pneumonia status (case-control) is directly caused by the presence or absence of a pathogen-caused infection in the lung. For controls, the lung is known to be sterile and has no infection. For a child clinically diagnosed with pneumonia, the pathogen causes the infection in a child's lung is the scientific target of interest. Among the candidate pathogens being tested, we assume only one is the primary cause. Extensions to multiple pathogens are straightforward. Because, for most cases, it is not possible to directly sample the lung, we do not know with certainty which pathogen infected the lung, so we seek to infer the infection status based upon a series of laboratory measurements of specimens from various body fluids and body sources. 

PERCH was originally designed with three sources of measurements relevant to the lung infection:  directly from the lung by lung aspirate; from blood culture; and from the nasopharyngeal cavity (by swab). Therefore, our model was designed to accommodate all three sources. As the study progressed, less than $1\%$ of cases had direct lung measurements and this sampled group was unrepresentative of all cases. The model and software here include all three sources of measurements for application to other etiology studies, but the analysis of the motivating PERCH data below uses only blood culture and nasopharyngeal swab data. 

The measurement error rates differ by type of measurement. Here, an error rate or \textit{epidemiologic} error rate is the probability of the pathogen's presence/absence in a specimen test given presence/absence of infection in the lung. For this application, it is convenient to categorize measures into three subgroups referred to as ``gold", ``silver", and ``bronze" standard measurements. A gold-standard (GS) measurement is assumed to have both perfect sensitivity and specificity. Lung aspirate data would have been gold-standard. A silver-standard (SS) measurement is assumed to have perfect specificity, but imperfect sensitivity. Culturing bacteria from blood samples (B-cX) is an example of silver-standard measurements in PERCH. Finally, bronze-standard (BrS) measurements are assumed to have imperfect sensitivity and specificity. Polymerase chain reaction (PCR) evaluation of bacteria and viruses from nasopharyngeal samples is an example.

In the PERCH study, both SS and BrS measurements are available for all cases. BrS, but not SS measures are available for controls. Our goal was to develop a statistical model that combines GS and SS measurements from cases, with BrS data from cases and controls to estimate the distribution of pathogens in the population of pneumonia cases, and the conditional probability that each of the $J$ pathogens is the primary cause of an individual child's pneumonia given her or his set of measurements. Even in applications where GS data is not available, a flexible modeling framework that can accommodate GS data is useful for both the evaluation of statistical information from BrS data (Section \ref{sec:simulation}) and the incorporation of GS data if it becomes available as measurement technology improves.

Latent class models (LCM) \citep{Goodman1974} have been successfully used to integrate multiple diagnostic tests or raters' assessments to estimate a binary latent status for all study subjects \citep{hui1980estimating,Qu1998,albert2001latent,albert2008estimating}. In the LCM framework, conditional distributions of measurements given latent status are specified. Then the marginal likelihood of the multivariate measurements are maximized as a function of the disease prevalence, sensitivities and specificities. This framework has also been extended to infer ordinal latent status \citep{wang2011evaluation}.

There are three salient features of the PERCH childhood pneumonia problem that require extension of the typical LCM approach. First, we have {\it partial} knowledge of the latent lung state for some subjects as a result of the case-control design. In the standard LCM approach, the study population comprises subjects with completely unknown class membership. In this study, controls are known to have no pathogen infecting the lung. Also, were gold standard measurements available from the lung for some cases, their latent variable would be directly observed. As the latent state is known for a non-trivial subset of the study population, we refer to this model as a partially-Latent Class Model or pLCM. 

Second, in most LCM applications, the number of observed measurements on a subject is much larger than the number of latent state categories. Here, the number of observations is of the same order as the number of categories that the latent status can assume. For example, if we consider only the PERCH study BrS data, we simultaneously observe the presence/absence of each member from a list of possible pathogens for each child. Even with additional control data, the larger number of latent categories of latent status leads to weak model identifiability as is discussed in more detail in Section \ref{sec:identifiability}.

Lastly, measurements with differing error rates (i.e. GS, SS, BrS) need to be integrated. Note that the modeling framework introduced here is general and can be applied to studies where multiple BrS measurements are available, each with a different set of error rates. Understanding the relative value of each level of measurements is important to optimally invest resources into data collection (number of subjects, type of samples) and laboratory assays. An important goal is therefore to estimate the relative information from each type of measurements about the population and individual etiology distributions. 

\cite{albert2008estimating} studied a model where some subjects are selected to verify their latent status (i.e. collect GS measurements) with the probability of verification either depending on the previous test results or being completely at random. They showed GS data can make model estimates more robust to model misspecifications. We further quantify how much GS data reduces the variance of model parameter estimates for design purposes. Also, they considered binary latent status and did not have available control data. Another related literature that uses both GS and BrS data is on verbal autopsy (VA) in the setting where no complete vital registry system is established in the community \citep{King2008}. Quite similar to the goal of inferring pneumonia etiology from lab measurements, the VA goal is to infer the cause of death from a pre-specified list by asking close family members questions about the presence/absence of several symptoms. \cite{King2008} proposed estimating the cause-of-death distribution in a community using data on dichotomous symptoms and GS data from the hospital where cause-of-death and symptoms are both recorded. However, their method involves nonparametric and requires a sizable sample of GS data, especially when the number of symptoms is large. In addition, a key difference between VA and most infectious disease etiology studies is that the VA studies are by definition case-only. 

Another approach previously used with case and control data is to perform logistic regression of case status on laboratory measurements and then to calculate point estimates of population attributable risks for each pathogen \citep{Bruzzi1985,Blackwelder2012}. This method does not account for imperfect laboratory measurements and cannot use GS or SS data if available. Also, the population attributable fraction method assigns zero etiology for the subset of pathogens that have estimated odds ratios smaller than $1$, without taking account of the statistical uncertainty for the odds ratio estimates.

In this paper, we define and apply a partially-latent class model (pLCM) to incorporate these three features: known infection status for controls; a large number of latent classes; and multiple types of measurements. We use a hierarchical Bayesian formulation to estimate: (1) the {\it population etiology distribution} or {\it etiology fraction} ---the frequency with which each pathogen ``causes" clinical pneumonia in the case population;  and (2) the {\it individual etiology probabilities}---the probabilities that a case is ``caused" by each of the candidate pathogens, given observed specimen measurements for that individual.

In Section \ref{sec:results}, to facilitate communications with scientists, we introduce graphical displays that put data, model assumptions, and results together. They enable the scientific investigators to better understand the various sources of evidence from data and their contribution to the final etiology estimates.

The remainder of this paper proceeds as follows. In section \ref{sec:models}, we formulate the pLCM and the Gibbs sampling algorithms for implementation. In Section \ref{sec:simulation}, we evaluate our method through simulations tailored for the childhood pneumonia etiology study. Section \ref{sec:results} presents the analysis of PERCH data. Lastly, Section \ref{sec:discussion} concludes with a discussion of results and limitations, a few natural extensions of the pLCM also motivated by the PERCH data, as well as future directions of research.

\section{A partially-latent class model for multiple indirect measurements}
\label{sec:models}
We develop pLCM to address two characteristics of the motivating pneumonia problem: (1) a partially-latent state variable because the pathogen infection status is known for controls but not cases; and (2) multiple categories of measurements with different error rates across classes. 
As shown in Figure \ref{fig:basicstructure}, let $I^L_i$, taking values in $\{0,1,2,...J\}$, represent the true state of child $i$'s lung ($i=1,...,N$) where $0$ represents no infection (control) and $I^L_i=j$, $ j=1,...,J$, represents the $j$th pathogen from a pre-specified cause-of-pneumonia list that is assumed to be exhaustive. $I^L_i$ is the scientific target of inference for individual diagnosis. Let $\bm{M}^S_i$ represent the $J\times 1$ vector of binary indicators of the presence/absence of each pathogen in the measurement at site $S$, where, in our childhood pneumonia etiology study $S$ can be nasopharyngeal (NP), blood (B), or lung (L). Let $\bm{m}_i^{S}$ be the actual observed values. In the following, we replace $S$ with BrS, SS, or GS, because they correspond to the measurement types at NP, B, and L, respectively.

Let $Y_i=y_i\in \{0,1\}$ represent the indicator of whether child $i$ is a healthy control or a clinically diagnosed case. Note $I^L_i=0$ given $Y_i=0$. To formalize the pLCM, we define three sets of parameters: 
\begin{itemize}
\item $\bm{\pi}=(\pi_1,...,\pi_J)'$ , the vector of compositional probabilities for each of $J$ pathogen causes, that is, $\text{Pr}(I_i^L =j \mid Y_i=1, \bm{\pi}), j= 1, ..., J$;
\item $\psi^S_j =\text{Pr}(M_{ij}^S =1|I_i^L=0)$, the false positive rate (FPR) for measurement $j$ ($j=1,...,J$) at site S. Note that The FPRs $\{\psi^S_j\}_{j=1}^J$ can be estimated from the control data at site S, because $I_i^L=0$ denotes that the $i$th subject has no infection in the lung, i.e. a control;
\item $\theta^S_j = \text{Pr}(M_{ij}^S =1|I_i^L=j)$, the true positive rate (TPR) for measurement $j$ at site $S$ for a person whose lung is infected by pathogen $j$, $j=1,...J$.
\end{itemize}
We further let $\bm{\psi}^S=(\psi_1^S,...,\psi_J^S)'$ and $\bm{\theta}^S=(\theta_1^S,...,\theta_J^S)'$. Using these definitions, we have FPR $\psi^{\GS}_j=0$ and TPR $\theta^{\GS}_j=1$ for GS measurements, so that $M^{\GS}_{ij}=1$ if and only if $I_i^L=j$, $j = 1, ..., J$ (perfect sensitivity and specificity). For SS measurements, FPR $\psi_j^{\SSs}=0$ so that $M^{\SSs}_j=0$ if $I_i^L\neq j$ (perfect specificity). 

We formalize the model likelihood for each type of measurement. We first describe the model for BrS measurement $\bm{M}^{\BrS}$ for a control or a case.
For control $i$, positive detection of the $j$th pathogen is a false positive representation of the non-infected lung. Therefore, we assume $M^{\BrS}_{ij} \mid \bm{\psi}^{\BrS} \sim \text{Bernoulli}(\psi_j^S), j=1,...,J$, with conditional independence, or equivalently,
\begin{eqnarray}
P^{0,\BrS}_{i} =
\text{Pr}(\bm{M}^{\BrS}_i= \bm{m}\mid \bm{\psi}^{\BrS})& = &  \prod_{j=1}^J\left(\psi_j^{\BrS}\right)^{m_j}\left(1-\psi_j^{\BrS}\right)^{1-m_j},
\label{eq:BrS_lkd_ctrl}
\end{eqnarray}
where $\bm{m}=\bm{m}_i^{\BrS}$.
For a case infected by pathogen $j$, the positive detection rate for the $j$th pathogen in BrS assays is $\theta^{\BrS}_j$. Since we assume a single cause for each case, detection of pathogens other than $j$ will be false positives with probability equal to FPR as in controls: $\psi_l^{\BrS}$, $l\neq j$. This nondifferential misclassification across the case and control populations is the essential assumption of the latent class approach because it allows us to borrow information from control BrS data to distinguish the true cause from background colonization. We further discuss it in the context of the pneumonia etiology problem in the final section. Then,
\begin{eqnarray}
\lefteqn{P^{1,\BrS}_{i} =\text{Pr}(\bm{M}^{\BrS}_{i}=\bm{m}\mid\bm{\pi}, \bm{\theta}^{\BrS}, \bm{\psi}^{\BrS})}\nonumber\\
 & = & \sum_{j=1}^J\pi_{j}\cdot \left(\theta_j^{\BrS}\right)^{m_j}\left(1-\theta^{\BrS}_j\right)^{1-m_j}\prod_{l\neq j}\left(\psi_l^{\BrS}\right)^{m_l}\left(1-\psi_l^{\BrS}\right)^{1-m_l}, \label{eq:casebrs_lkd}
\label{eq:BrS_lkd_case}
\end{eqnarray}
is the likelihood contributed by BrS measurements from case $i$, where $\bm{m}=\bm{m}_{i}^{\BrS}$.

Similarly, likelihood contribution from case $i$'s SS measurements can be written as
\begin{eqnarray}
P^{1,\text{SS}}_{i}  = 
\text{Pr}(\bm{M}^{\SSs}_{i}=\bm{m}\mid \bm{\pi}, \bm{\theta}^{\SSs})= \sum_{j=1}^{J'}\pi_{j}\cdot \left(\theta_j^{\SSs}\right)^{m_j}(1-\theta^{\SSs}_j)^{1-m_j}\mathbf{1}_{\left\{\sum_{l=1}^{J'}m_l\leq 1\right\}},
\label{eq:GnS_lkd}
\end{eqnarray}
for $\bm{m}=\bm{m}_{i}^{\SSs}$, noting the perfect specificity of SS measurements, where $J'\leq J$ represents the number of actual SS measurements on each case, and $\bm{\theta}^{\SSs} = \left(\theta_1^{\SSs},...\theta_{J'}^{\SSs}\right)$. SS measurements only test for a subset of all $J$ pathogens, e.g., blood culture only detects bacteria and $J'$ is the number of bacteria that are potential causes. Finally, for completeness, GS measurement is assumed to follow a multinomial distribution with likelihood: 
\begin{eqnarray}
P^{1,\text{GS}}_{i}=\text{Pr}\left(M^{\GS}_{i}  =  \bm{m}\mid \bm{\pi}\right) 
& = & \prod_{j=1}^J \pi_j^{\mathbf{1}\left\{m_j=1\right\}}\mathbf{1}_{\{\sum_j{m_{j}=1}\}},\label{eq:GS_lkd}
\end{eqnarray}
where $\bm{m}=\bm{m}_{i}^{\GS}$, and $\mathbf{1}_{\{\cdot\}}$ is the indicator function and equals one if the statement in $\{\cdot\}$ is true; otherwise, zero.

Let $\delta_i$ be the binary indicator of a case $i$ having GS measurements; it equals $1$ if the case has available GS data and $0$ otherwise.  Combining likelihood components (\ref{eq:BrS_lkd_ctrl})---(\ref{eq:GS_lkd}), the total model likelihood for BrS, SS, and GS data across independent cases and controls is
\begin{eqnarray}
L(\bm{\gamma}; \mathcal{D}) & = &
\prod_{i: Y_i=0}P_{i}^{0,\BrS}
\prod_{i:Y_{i}=1, \delta_{i}=1}P_{i}^{1,\BrS}
\cdot P_{i}^{1,\SSs}
\cdot P_{i}^{1,\GS}
\prod_{i:Y_{i}=1, \delta_{i}=0}P_{i}^{1,\BrS}
\cdot P_{i}^{1,\SSs}
,\label{eq:model}
\end{eqnarray}
where $\bm{\gamma}=(\bm{\theta}^{\BrS},\bm{\psi}^{\BrS}, \bm{\theta}^{\SSs},\bm{\pi})'$ stacks all unknown parameters, and data $\mathcal{D}$ is
$$\left\{\left\{\bm{m}_{i}^{\BrS}\right\}_{i: Y_i=0}\right\}\cup \left\{\left\{\bm{m}_{i}^{\BrS},\bm{m}_{i}^{\GS},\bm{m}_{i}^{\SSs}\right\}_{i: Y_{i}=1, \delta_{i}=1}\right\} \cup \left\{\left\{\bm{m}_{i^{''}}^{\BrS},\bm{m}_{i^{''}}^{\SSs}\right\}_{i^{''}: Y_{i^{''}}=1,\delta_{i^{''}}=0}\right\}$$
collects all the available measurements on study subjects. Our primary statistical goal is to estimate the posterior distribution of the population etiology distribution $\bm{\pi}$, and to obtain individual etiology ($I^L_{*}$) prediction given a case's measurements.

To enable Bayesian inference, prior distributions on model parameters are specified as follows:
$\bm{\pi}  \sim  \text{Dirichlet}(a_1,\dots,a_{J})$, 
$\psi_j^{\BrS} \sim \text{Beta}(b_{1j},b_{2j})$,
$\theta_j^{\BrS} \sim  \text{Beta}(c_{1j},c_{2j}), j=1,...,J$, and 
$\theta_j^{\SSs} \sim  \text{Beta}(d_{1j},d_{2j})$, $j=1,...,J'$. Hyperparameters for etiology prior, $a_1, ..., a_J$, are usually $1$s to denote equal and non-informative prior weights for each pathogen if expert prior knowledge is unavailable. The FPR for the $j$th pathogen, $\psi_j^{\BrS}$, generally can be well estimated from control data, thus $b_{1j}=b_{2j}=1$ is the default choice. For TPR parameters $\theta_j^{\BrS}$ and $\theta_j^{\SSs}$, if prior knowledge on TPRs is available, we choose $(c_{1j},c_{2j})$ so that the $2.5\%$ and $97.5\%$ quantiles of Beta distribution with parameter $(c_{1j},c_{2j})$ match the prior minimum and maximum TPR values elicited from pneumonia experts . Otherwise, we use default value $1$s for the Beta hyperparameters. Similarly we choose values of $(d_{1j},d_{2j})$ either by prior knowledge or default values of $1$. We finally assume prior independence of the parameters as $[\bm{\gamma}]=[\bm{\pi}][\bm{\psi}^{\BrS}][\bm{\theta}^{\BrS}][\bm{\theta}^{\SSs}]$, where $[A]$ represents the distribution of random variable or vector $A$. These priors represent a balance between explicit prior knowledge about measurement error rates and the desire to be as objective as possible for a particular study. As described in the next section, the identifiability constraints on the pLCM require specifying a reasonable subset of parameter values to identify parameters of greatest scientific interest. 

\subsection{Model identifiability}
\label{sec:identifiability}
Potential non-identifiability of LCM parameters is well-known \citep{Goodman1974}. For example, an LCM with four observed binary indicators and three latent classes is not identifiable despite providing $15$ degree-of-freedom to estimate $14$ parameters \citep{Goodman1974}. In principle, the Bayesian framework avoids the non-identifiability problem in LCMs by incorporating prior information about unidentified parameter subspaces (e.g., \citet{garrett2000latent}). Many authors point out that the posterior variance for non-identifiable parameters does not decrease to zero as sample size approaches infinity (e.g., \cite{Kadane1974, Gustafson2001, Gustafson2005}). Even when data are not fully informative about a parameter, an identified set of parameter values consistent with the observed data shall, can nevertheless, be valuable in a complex scientific investigation \citep{gustafson2009limits} like PERCH.

When GS data is available, the pLCM is identifiable; when it is not, the two sets of parameters, $\bm{\pi}$ and $\{\theta_j^{\BrS}\}_{j=1}^J$ are not both identified and prior knowledge must be incorporated. Here we restrict attention to the scenario with only BrS data for simplicity but similar arguments pertain to the BrS + SS scenario. The problem can be understood from the form of the positive measurement rates for pathogens among cases. In the pLCM likelihood for the BrS data (only retaining components in (\ref{eq:model}) with superscripts $\BrS$), the positive
rate for pathogen $j$ is a convex combination of the TPR and FPR: 
\begin{equation}
\text{Pr}\left(M^{\BrS}_{ij}=1\mid \pi_j,\theta_j^{\BrS},\psi_j^{\BrS}\right)=\pi_j\theta^{\BrS}_j+(1-\pi_j)\psi^{\BrS}_j,\label{eq:convex_comb}
\end{equation} 
where the left-hand side of the above equation can be estimated by the observed positive rate of pathogen $j$ among cases. Although the control data provide $\psi_j^{\BrS}$ estimates, the two parameters, $\pi_j$ and $\theta_j^{\BrS}$, are not both identified. GS data, if available, identifies $\pi_j$ and resolves the lack of identifiability. Otherwise, we need to incorporate prior scientific information on one of them, usually the TPR ($\theta_j^{\BrS}$). In PERCH, prior knowledge about $\theta_j^{\BrS}$ is obtained from infectious disease and laboratory experts \citep{Murdoch2012} based upon vaccine probe studies \citep{cutts2005efficacy,Madhi15052005}. If the observed case positive rate is much higher than the rate in controls ($\psi_j^{\BrS}$), only large values of TPR ($\theta_j^{\BrS}$) are supported by the data making etiology estimation more precise (Section \ref{sec:ind.pred}).

The full model identification can be generally characterized by inspecting the Jacobian matrix of the transformation $F$ from model parameters $\bm{\gamma}$ to the distribution $\bm{p}$ of the observables, $\bm{p}  = F(\bm{\gamma})$. Let $\bm{\gamma}=(\bm{\theta}^{\BrS},\bm{\psi}^{\BrS}, \pi_1,...,\pi_{J-1})'$ represent the $3J-1$-dimensional unconstrained model parameters. The pLCM defines the transformation $(\bm{p}_1,\bm{p}_0)'=F(\bm{\gamma})$, where $\bm{p}_1$ and $\bm{p}_0$ are the two contingency probability distributions for the BrS measurements in the case and control populations, each with dimension $2^J-1$. It can be shown that the Jacobian matrix has $J-1$ of its singular values being zero, which means model parameters $\bm{\gamma}$ are not fully identified from the data. The FPRs ($\psi_j^{\BrS}, j=1,...,J$) in pLCM are, however, identifiable parameters that can be estimated from control data. Therefore, pLCM is termed partially identifiable \citep{Jones2010}.

\subsection{Parameter estimation and individual etiology prediction}
\label{sec:ind.pred}

The parameters in likelihood (\ref{eq:model}) include the population etiology distribution ($\bm{\pi}$), TPRs and FPRs for BrS measurements ($\bm{\psi}^{\BrS}$ and $\bm{\theta}^{\BrS}$),  and TPRs for SS measurements ($\bm{\theta}^{\SSs}$). The posterior distribution of these parameters can be estimated by constructing approximating samples from the joint posterior via a Markov chain Monte Carlo (MCMC) Gibbs sampler. The full conditional distributions for the Gibbs sampler are detailed in Section A of the supplementary material.

We develop a Gibbs sampler with two essential steps: 
\begin{enumerate}
\item[(i)] Multinomial sampling of lung infection state among cases:\\
$I^L_{i}\mid \bm{\pi}, Y_{i}=1\sim \text{Multinomial}(\bm{\pi})$;
\item[(ii)] Measurement stage given lung infection state: 

$M^{\BrS}_{ij}\mid I^L_{i}, \bm{\theta}^{\BrS}, \bm{\psi}^{\BrS} \sim\text{Bernoulli}\left(\mathbf{1}_{\{I^L_{i}=j\}}\theta^{\BrS}_j+\left(1-\mathbf{1}_{\{I^L_{i}=j\}}\right)\psi^{\BrS}_j\right), j = 1,...,J$, conditionally independent.
\end{enumerate}

This is readily implemented using freely available software \verb"WinBUGS 1.4". In the application below, convergence was monitored using auto-correlations, kernel density plots, and Brooks-Gelman-Rubin statistics \citep{brooks1998general} of the MCMC chains. The statistical results below are based on $10,000$ iterations of burn-in followed by $50,000$ production samples from each of three parallel chains.  

The Bayesian framework naturally allows individual within-sample classification (infection diagnosis) and out-of-sample prediction. This section describes how we calculate the etiology probabilities for an individual with measurements $\bm{m}_{*}$. We focus on the more challenging inference scenario when only BrS data are available; the general case follows directly. 

The within-sample classification for case $i$ is based on the posterior distribution of latent indicators given the observed data, i.e. $\text{Pr}(I^L_{i}=j \mid \mathcal{D})$, $j=1,...,J$, which can be obtained by averaging along the cause indicator ($I^L_{i}$) chain from MCMC samples. For a case with new BrS measurements $\bm{m}_*$, we have
\begin{eqnarray}
\Pr(I^L_{i}=j\mid\bm{m}_*, \mathcal{D}) & = &\int \Pr(I^L_{i}=j\mid \bm{m}_*,\bm{\gamma})\Pr(\bm{\gamma} \mid \bm{m}_*,\mathcal{D}) \mathrm{d}\bm{\gamma},  j=1,...J,\label{eq:outofsamplepred}
\end{eqnarray}
where the second factor in the integrand can be approximated by the posterior distribution given current data, i.e., $\text{Pr}(\bm{\gamma} \mid  \mathcal{D})$. For the first term in the integrand, we explicitly obtain the model-based, one-sample conditional posterior distribution, $\text{Pr}(I^L_{i}=j \mid \bm{m}_*,\bm{\gamma}) = \pi_j \ell_j(\bm{m}_*; \bm{\gamma})\bigg /\sum_{m} \pi_{\bm{r}_m}\ell_m(\bm{m}_*; \bm{\gamma})$, $j=1,...,J$, where 
$$\ell_m(\bm{m}_*; \bm{\gamma})=\left(\theta_j^{\BrS}\right)^{m_{*j}}\left(1-\theta^{\BrS}_j\right)^{1-m_{*j}}\prod_{l\neq j}\left(\psi_l^{\BrS}\right)^{m_{*l}}\left(1-\psi_l^{\BrS}\right)^{1-m_{*l}}$$
 is the $m$th mixture component likelihood function evaluated at $\bm{m}_*$. The log relative probability of $I^L_i=j$ versus $I^L_i=l$ is 
\begin{eqnarray*}
\lefteqn{R_{jl}=\log \left(\frac{\pi_j}{\pi_l}\right)+
\log \left \{\left(\frac{\theta^{\BrS}_j}{\psi^{\BrS}_j}\right)^{m_{*j}}\left(\frac{1-\theta^{\BrS}_j}{1-\psi^{BrS}_j}\right)^{1-m_{*j}}\right \}}\\
&&\quad\quad\quad\quad\quad\quad\quad\quad\quad+\log\left\{\left(\frac{\psi^{\BrS}_l}{\theta^{\BrS}_l}\right)^{m_{*l}}\left(\frac{1-\psi^{\BrS}_l}{1-\theta^{\BrS}_l}\right)^{1-m_{*l}} \right \}.
\end{eqnarray*}
The form of $R_{jl}$ informs us about what is required for correct diagnosis of an individual. Suppose $I^L_i=j$, then averaging over $\bm{m}_{*}$,  we have $E[R_{jl}]= \log\left({\pi_j}/{\pi_l}\right)+I(\theta^{\BrS}_j; \psi^{\BrS}_j)+I(\psi^{\BrS}_l;\theta^{\BrS}_l)$, where $I(v_1;v_2)=v_1\log(v_1/v_2)+(1-v_1)\log\left((1-v_1)/(1-v_2)\right)$ is the information divergence \citep{kullback2012information} that represents the expected amount of information in $m_{*j}\sim \text{Bernoulli}(v_1)$ for discriminating against $m_{*j}\sim \text{Bernoulli}(v_2)$. If $v_1=v_2$, then $I(v_1;v_2)=0$. The form of $E[R_{jl}] $ shows that there is only additional information from BrS data about an individual's etiology in the person's data when there is a difference between $\theta_j^{\BrS}$ and $\psi_j^{\BrS}$, $j=1,...,J$. 

Following equation (\ref{eq:outofsamplepred}), we average $\text{Pr}(I^L_{i}=j \mid \bm{m}_*,\bm{\gamma})$ over MCMC iterations to obtain individual prediction for the $j$th pathogen, $\hat{p}_{ij}$, with $\bm{\gamma}$ replaced by its simulated values $\bm{\gamma}^*$ at each iteration. Repeating for $j=1,...,J$, we obtain a $J\time 1$ probability vector, $\hat{\bm{p}}_{i}=(\hat{p}_{i1},...,\hat{p}_{iJ})'$, that sums to one.  This scheme is especially useful when a newly examined case has a BrS measurement pattern not observed in $\mathcal{D}$, which often occurs when $J$ is large. The final decisions regarding which pathogen to treat can then be based upon $\hat{\bm{p}}_{i}$. In particular, the pathogen with largest posterior value might be selected. It is Bayes optimal under mean misclassification loss. Individual etiology predictions described here generalize the positive/negative predictive value (PPV/NPV) from single to multivariate binary measurements and can aid diagnosis of case subjects under other user-specified misclassification loss functions. 
 
\section{Simulation for three pathogens case with GS and BrS data}
\label{sec:simulation}


To demonstrate the utility of the pLCM for studies like PERCH, we simulate BrS data sets with $500$ cases and $500$ controls for three pathogens, A, B, and C using known pLCM specifications. We focus on three states to facilitate viewing of the $\bm{\pi}$ estimates and individual predictions in the 3-dimensional simplex $\mathcal{S}^2$. We use the ternary diagram \citep{Aitchison1986} representation where the vector $\bm{\pi}=(\pi_A,\pi_B,\pi_C)'$ is encoded as a point with each component being the perpendicular distance to one of the three sides. The parameters involved are fixed at  $\text{TPR}=\bm{\theta}=(\theta_A,\theta_B,\theta_C)'=(0.9,0.9,0.9)'$,  $\text{FPR}=\bm{\psi}=(\psi_A,\psi_B,\psi_C)'=(0.6,0.02,0.05)'$, and $\bm{\pi}=(\pi_A,\pi_B,\pi_C)'=(0.67,0.26,0.07)'$. We focus on BrS and GS data here and have dropped the ``$\BrS$" superscript on the parameters for simplicity. We further let the fraction of cases with GS measurements ($\Delta$) be either $1\%$ as in PERCH or $10\%$. Although GS measurements are rare in the PERCH study, we investigate a large range of $\Delta$ to understand in general how much statistical information is contained in BrS measurements relative to GS measurements.

\begin{figure}[!p]
\begin{subfigure}{.5\textwidth}
  \centering
  \includegraphics[width=\linewidth]{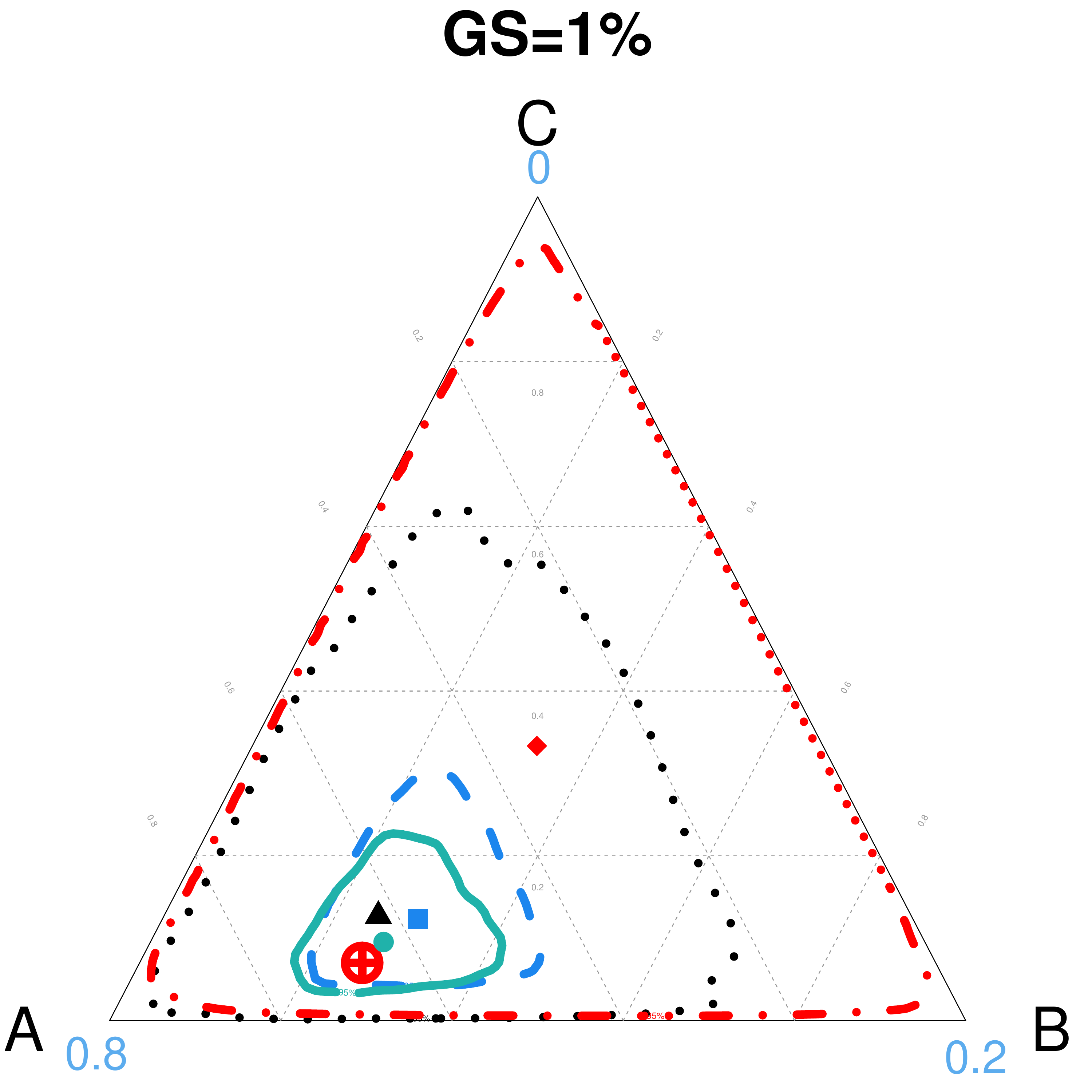}
  \caption[1a]{}
  \label{fig:BSvalue_a}
\end{subfigure}
\begin{subfigure}{.5\textwidth}
  \centering
  \includegraphics[width=\linewidth]{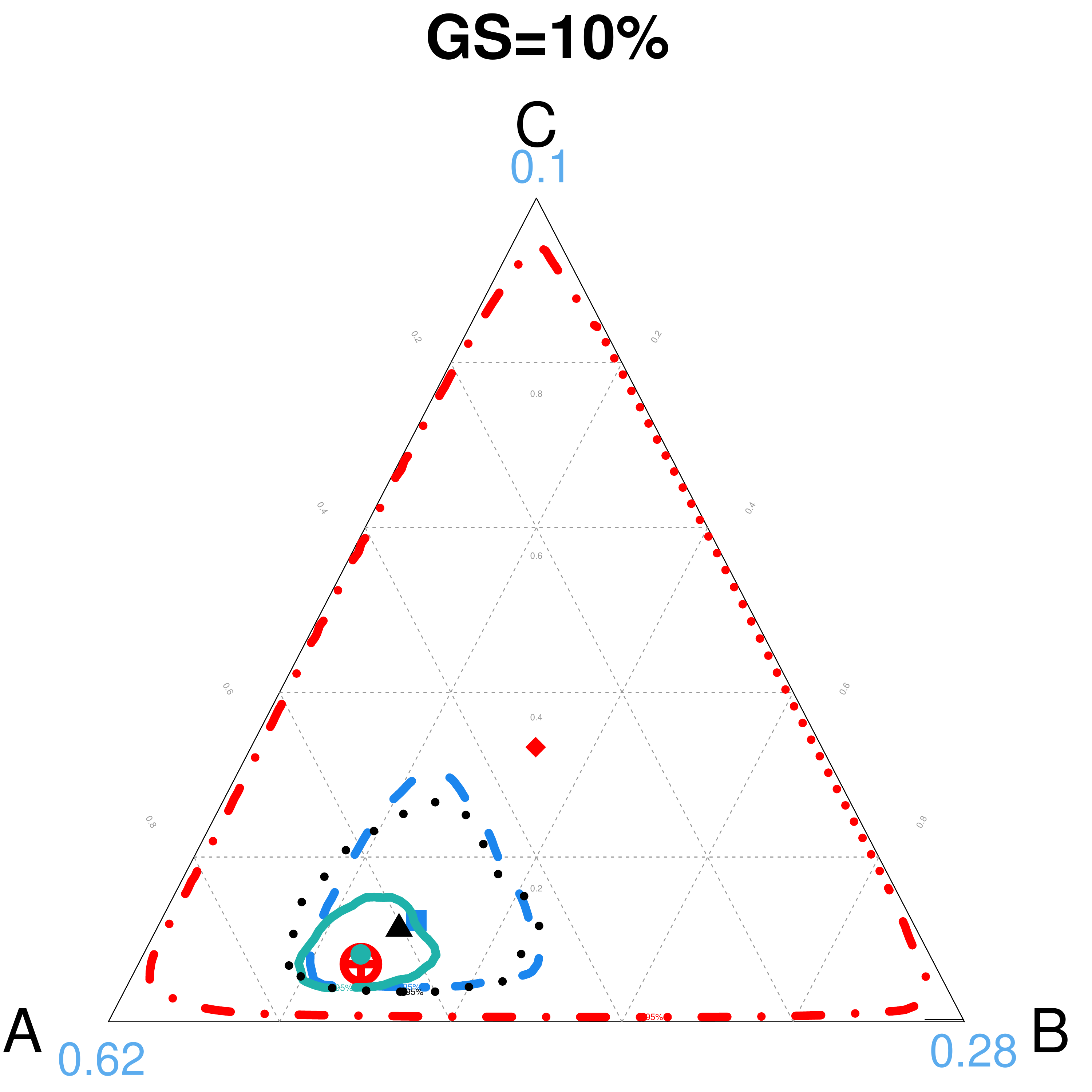}
  \caption{}
  \label{fig:BSvalue_b}
\end{subfigure}
\begin{subfigure}{.5\textwidth}
  \centering
  \includegraphics[width=\linewidth]{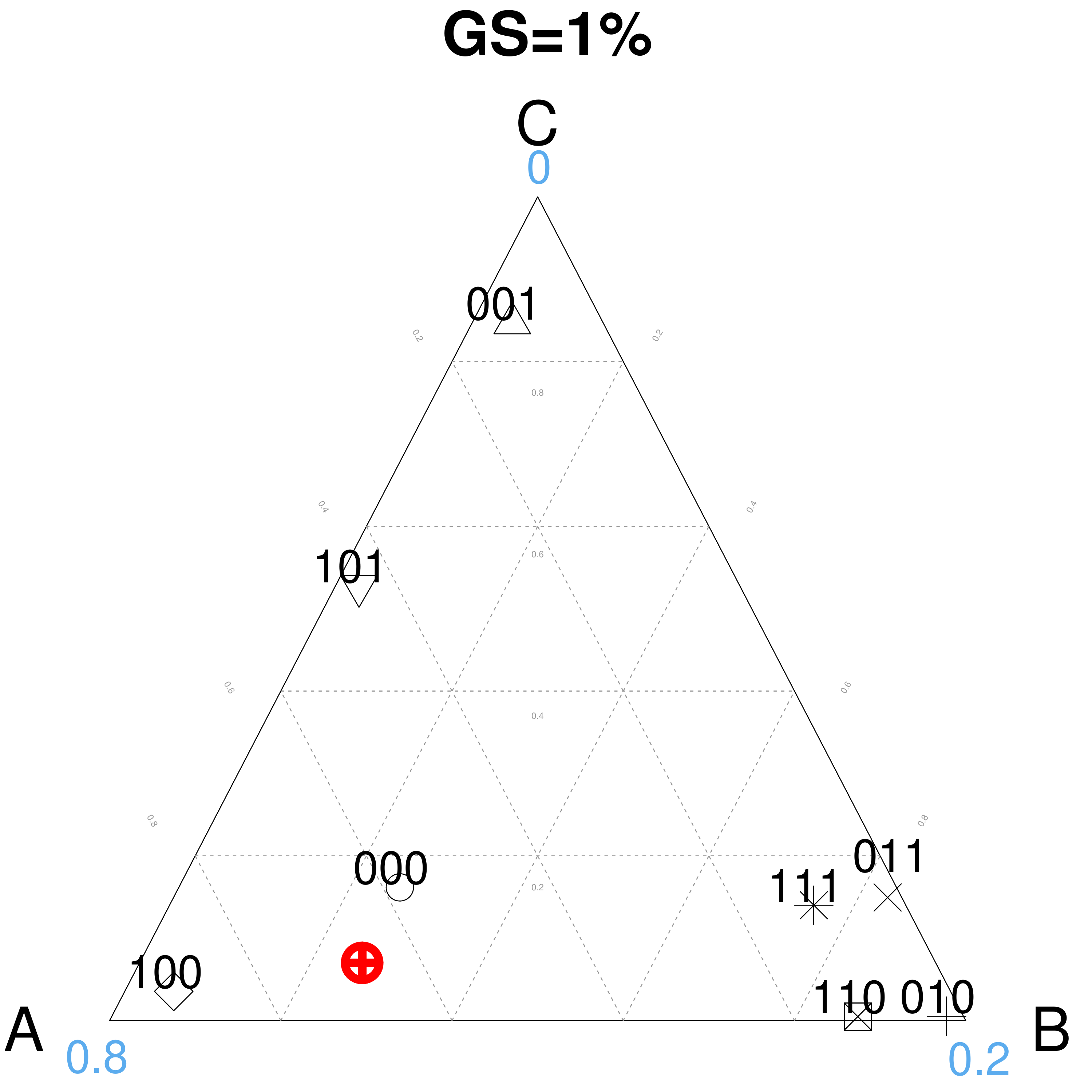}
  \caption[1a]{}
  \label{fig:BSvalue_a_ind}
\end{subfigure}
\begin{subfigure}{.5\textwidth}
  \centering
  \includegraphics[width=\linewidth]{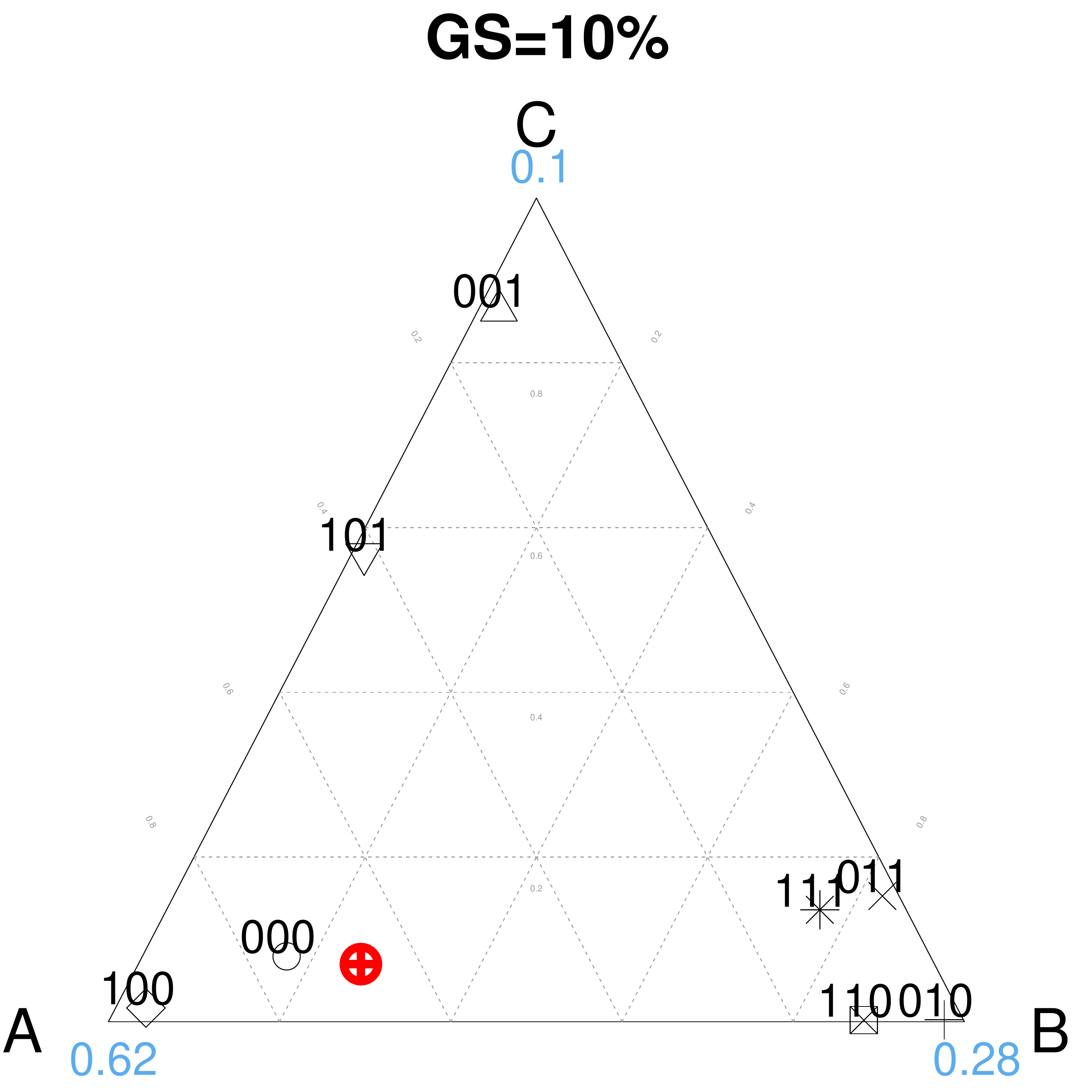}
  \caption{}
  \label{fig:BSvalue_b_ind}
\end{subfigure}
\caption[Population (top) and individual (bottom) etiology estimates for a single sample with $500$ cases and $500$ controls.]{Population (top) and individual (bottom) etiology estimations for a single sample with $500$ cases and $500$ controls with true $\bm{\pi}=(0.67,0.26,0.07)'$ and either $1\%(N=5)$ or $10\%(N=50)$ GS data on cases. In (a) or (b), \textit{red circled plus} shows the true population etiology distribution $\bm{\pi}$. The closed curves are $95\%$ credible regions for analysis using BrS data only (\textit{blue dashed lines} ``- - -"), BrS+GS data (\textit{light green solid lines} ``---"), GS data only (textit{black dotted lines} ``$\cdots$"); \textit{Solid square/dot/triangle} are the corresponding posterior means of $\bm{\pi}$; The $95\%$ highest density region of uniform prior distribution is also visualized by red ``$\cdot - \cdot -$" for comparison. In (c) or (d), $8 (=2^3)$ BrS measurement patterns and predictions for individual children are shown with measurement patterns attached. The numbers at the vertices show empirical frequencies of GS measurements.}
\label{fig:BSvalue.simulation}
\end{figure}

For any given data set, three distinct subsets of the data can be used: BrS-only, GS-only, and BrS+GS, each producing its posterior mean of $\bm{\pi}$, and $95\%$ credible region (Bayesian confidence region) by transformed Gaussian kernel density estimator for compositional data \citep{Chacon2011}. To study the relative importance of the GS and BrS data, the primary quantity of interest in the simulations is the relative sizes of the credible regions for each data mix. Here, we use uniform priors on $\bm{\theta}$, $\bm{\psi}$, and $\text{Dirichlet}(1,...,1)$ prior for $\bm{\pi}$. The results are shown in Figure \ref{fig:BSvalue.simulation}.

First, in Figures \ref{fig:BSvalue_a} ($1\%$ GS) and \ref{fig:BSvalue_b} ($10\%$ GS), each region covers the true etiology $\bm{\pi}$. In data not shown here, the nominal $95\%$ credible regions covers slightly more than $95\%$ of $200$ simulations. Credible regions narrow in on the truth as we combine BrS and GS data, and as the fraction of subjects with GS data ($\Delta$) increases. Also, the posterior mean from the BrS+GS analysis is an optimal balance of information contained in  the GS and BrS data.

Using the same simulated data sets, Figures \ref{fig:BSvalue_a_ind} and \ref{fig:BSvalue_b_ind} also show individual etiology predictions for each of the $8 (=2^3)$ possible BrS measurements $(m_A,m_B,m_C)', m_j=0,1$, obtained by the methods from Section \ref{sec:ind.pred}. Consider the example of a newly enrolled case without GS data and with no pathogen observed in her BrS data: $\bm{m}=(0,0,0)'$. Suppose she is part of a case population with $10\%$ GS data. In the case illustrated in Figure \ref{fig:BSvalue_b_ind}, her posterior predictive distribution has highest posterior probability ($0.76$) on pathogen A reflecting two competing forces: the FPRs that describe background colonization (colonization among the controls) and the population etiology distribution. Given other parameters, $\bm{m}=(0,0,0)'$ gives the smallest likelihood for $I^L_i=A$ because of its high background colonization rate (FPR $\psi_A=0.6$). However, prior to observing $(0,0,0)'$, $\pi_A$ is well estimated to be much larger than $\pi_B$ and $\pi_C$. Therefore the posterior distribution for this case is heavily weighted towards pathogen A. 

Because it is rare to observe pathogen $B$ in a case whose pneumonia is not caused by B, for a case with observation $(1,1,1)'$, the prediction favors B. Although B is not the most prevalent cause among cases, the presence of B in the BrS measurements gives the largest likelihood when $I_i^L=B$. For any measurement pattern with a single positive, the case is always classified into that category in this example.

Most predictions are stable with increasing gold-standard percentage, $\Delta$. Only $000$ cases have predictions that move from near the center to the corner of A. This is mainly because that TPR $\bm{\theta}$ and etiology fractions $\bm{\pi}$ are not as precisely estimated in GS-scarce scenarios relative to GS-abundant ones. Averaging over a wider range of $\bm{\theta}$ and $\bm{\pi}$ produces $000$ case predictions that are ambiguous, i.e. near the center. As $\Delta$ increases, parameters are well estimated, and precise predictions result.

\section{Analysis of PERCH data}
\label{sec:results}
The Pneumonia Etiology Research for Child Health (PERCH) study is an on-going standardized
and comprehensive evaluation of etiologic agents causing severe and very severe
pneumonia among hospitalized children aged $1$-$59$ months in seven low and middle income countries \citep{Levine2012}. The study sites include countries with a significant burden of childhood pneumonia and a range of epidemiologic characteristics. PERCH is a case-control study that has enrolled over $4,000$ patients hospitalized for severe or very severe pneumonia and over $5,000$ controls selected randomly from the community frequency-matched on age in each month. More details about the PERCH design are available in \cite{Deloria2012}. 

To analyze PERCH data with the pLCM model, we have focused on preliminary data from one site with good availability of both SS and BrS laboratory results (no missingness). Final analyses of all $7$ countries will be reported elsewhere upon study completion. Included in the current analysis are BrS data (nasopharyngeal specimen with PCR detection of pathogens) for $432$ cases and $479$ frequency-matched controls on $11$ species of pathogens ($7$ viruses and $4$ bacteria; representing a subset of pathogens evaluated; their abbreviations shown on the right margin in Figure \ref{fig:site02GAM}, and full names in Section B of the supplementary material), and SS data (blood culture results) on the $4$ bacteria for only the cases. 

In PERCH, prior scientific knowledge of measurement error rates is incorporated into the analysis. Based upon microbiology studies \citep{Murdoch2012}, the PERCH investigators selected priors for the TPRs of our BrS measurements, $\theta_j^{\BrS}$, in the range of $50\%-100\%$ for viruses and $0-100\%$ for bacteria. Priors for the SS TPRs were based on observations from vaccine probe studies---randomized clinical trials of pathogen-specific vaccines where the total number of clinical pneumonia cases prevented by the vaccine is much larger than the few SS laboratory-confirmed cases prevented. Comparing the total preventable disease burden to the number of blood culture (SS) positive cases prevented provides information about the TPR of the bacterial blood culture measurements, $\theta_j^{\SSs},j=1,...,4$. Our analysis used the range $5-15\%$ for the SS TPRs of the four bacteria consistent with the vaccine probe studies \citep{cutts2005efficacy, Madhi15052005}. We set Beta priors that match these ranges (Section \ref{sec:models}) and assumed Dirichlet($1,...,1$) prior on etiology fractions $\bm{\pi}$.

In latent variable models like the pLCM, key variables are not directly observed. It is therefore essential to picture the model inputs and outputs side-by-side to better understand the analysis. In this spirit, Figure \ref{fig:site02GAM} displays for each of the 11 pathogens, a summary of the BrS and SS data in the left two columns, along with some of the intermediate model results; and the prior and posterior distributions for the etiology fractions on the right (rows ordered by posterior means).  The observed BrS rates (with $95\%$ confidence intervals) for cases and controls are shown on the far left with solid dots. The conditional odds ratio contrasting the case and control rates given the other pathogens is listed with $95\%$ confidence interval in the box to the right of the BrS data summary. Below the case and control observed rates is a horizontal line with a triangle. From left to right, the line starts at the estimated false positive rate (FPR, $\hat{\psi}_j^{\BrS}$) and ends at the estimated true positive rate (TPR, $\hat{\theta}_j^{\BrS}$), both obtained from the model. Below the TPR are two boxplots summarizing its posterior (top) and prior (bottom) distributions for that pathogen. These box plots show how the prior assumption influences the TPR estimate as expected given the identifiability constraints discussed in Section \ref{sec:identifiability}. The triangle on the line is the model estimate of the case rate to compare to the observed value above it. As discussed in Section \ref{sec:identifiability}, the model-based case rate is a linear combination of the FPR and TPR with mixing fraction equal to the estimated etiology fraction. Therefore, the location of the triangle, expressed as a fraction of the distance from the FPR to the TPR, is the model-based point estimate of the etiologic fraction for each pathogen. The SS data are shown in a similar fashion to the right of the BrS data. By definition, the FPR is $0.0\%$ for SS measures and there is no control data. The observed rate for the cases is shown with its $95\%$ confidence interval. The estimated SS TPR ($\hat{\theta}^{\SSs}_j$) with prior and posterior distributions is shown as for the BrS data, except that we plot $95\%$ and $50\%$ credible intervals for SS TPR above its prior $95\%$ and $50\%$ intervals.

On the right side of the display are the marginal posterior and prior distributions of the etiologic fraction for each pathogen. We appropriately normalized each density to match the height of the prior and posterior curves. The posterior mean, $50\%$ and $95\%$ credible intervals are shown above the density.

Figure \ref{fig:site02GAM} shows that respiratory syncytial virus (RSV), \textit{Streptococcus pneumoniae} (PNEU), rhinovirus (RHINO), and human metapneumovirus (HMPV$\_$A$\_$B) occupy the greatest fractions of the etiology distribution, from $15\%$ to $30\%$ each. That RSV has the largest estimated mean etiology fraction reflects the large discrepancy between case and control positive rates in the BrS data: $25.1\%$ versus $0.8\%$ (marginal odds ratio $38.5$ ($95\%$\text{CI} $(18.0,128.7)$ ). RHINO has case and control rates that are close to each other, yet its estimated mean etiology fraction is $16.7\%$. This is because the model considers the joint distribution of the pathogens, not the marginal rates. The conditional odds ratio of case status with RHINO given all the other pathogen measures is estimated to be $1.5$ $(1.1,2.1)$ as in contrast to the marginal odds ratio close to $1$ $(0.8,1.3)$. 

As discussed in Section \ref{sec:identifiability}, the data alone cannot precisely estimate both the etiologic fractions and TPRs absent prior knowledge. This is evidenced by comparing the prior and posterior distributions for the TPRs in the BrS boxes for some pathogens like HMPV$\_$A$\_$B and PARA1 (i.e. left hand column of Figure \ref{fig:site02GAM}). The posteriors are similar to their priors indicating little else about TPR is learned from the data. The posteriors for some pathogens making up $\bm{\pi}$ (i.e. shown in the right hand column of Figure \ref{fig:site02GAM}) are likely to be sensitive to the prior specifications of the TPRs. 

We performed sensitivity analyses using multiple sets of priors for the TPRs. At one extreme, we ignored background scientific knowledge and let the priors on the FPR and TPR be uniform for both the BrS and SS data. Ignoring prior knowledge about error rates lowers the etiology estimates of the bacteria PNEU and \textit{Staphylococcus aureus} (SAUR). The substantial reduction in the etiology fraction for PNEU, for example, is a result of the difference in the TPR prior for the SS measurements. In the original analysis (Figure \ref{fig:site02GAM}), the informative prior on the SS sensitivity (TPR) places $95\%$ mass between $5-15\%$. Hence the model assumes almost $90\%$ of the PNEU infections are being missed in the SS sampling. When a uniform prior is substituted, the fraction assumed missed is greatly reduced. For RSV, its posterior mean etiology fraction is stable ($29.4\%$ to $30.0\%$). The etiology estimates for other pathogens are fairly stable, with changes in posterior means between $-2.3\%$ and $3.4\%$. 


Under the original priors for TPR, PARA1 has an estimated etiologic fraction of $6.4\%$, even though it has conditional odds ratio $5.9~(2.6,15.0)$. In general, pathogens with larger conditional odds ratios have larger etiology fraction estimates. But a pathogen also needs a reasonably high observed case positive rate to be allocated a high etiology fraction. The posterior etiology fraction estimate of $6.4\%$ for PARA1 results because the prior for the TPR takes values in the range of $50-99\%$. By Equation (\ref{eq:convex_comb}), the TPR weight in the convex combination with FPR (around $1.5\%$) has to be very small to explain the small observed case rate $5.6\%$. When a uniform prior is placed on TPR instead, the PARA1 etiology fraction increases to $9.4\%$ with a wider $95\%$ credible interval.

We believe that RHINO's etiologic fraction may be inflated as a result of its negative association with RSV among cases. Under the conditional independence assumption of the pLCM, this dependence can only be explained by multinomial correlation among the latent cause indicators: $I^L_i=\text{\footnotesize RSV}$ versus $I^L_i=\text{\footnotesize RHINO}$ that is $-\pi_{\text{\footnotesize RSV}}\pi_{\text{\footnotesize RHINO}}$. There is strong evidence that RSV is a common cause with a stable estimate $\hat{\pi}_{\text{\footnotesize RSV}}$ around $30\%$. The strong negative association in the cases' measurements between RHINO and RSV therefore is being explained by a larger etiologic fraction estimate $\hat{\pi}_{\text{\footnotesize RHINO}}$ relative to other pathogens that have less or no association with RSV among the cases. The conditional independence assumption is leveraging information from the associations between pathogens in estimation of the etiologic fractions. If true, this issue can be addressed by extending the pLCM to allow for alternate sources of correlation among the measurements, for example, competition among pathogens within the NP space.

We have checked the model in two ways by comparing the characteristics of the observed measurements joint distribution with the same characteristic for the distribution of data of the same size generated by the model. By generating the new data characteristics at every iteration of the MCMC chain, we can obtain the posterior predictive distribution by integrating over the posterior distribution of the parameters  \citep{garrett2000latent}. 

Among the cases, the $95\%$ predictive interval includes the observed values in all but two of the BrS patterns and even there the fits are reasonable. Among the controls, there is evidence of lack of fit for the most common BrS pattern with only PNEU and HINF (Figure S1 in supplementary materials). There are fewer cases with this pattern observed than predicted under the pLCM. This lack of fit is likely due to associations of pathogen measurements in control subjects. Note that the FPR estimates remain consistent regardless of such correlation as the number of controls increases, however posterior variances for them may be underestimated.

A second model-checking procedure is for the conditional independence assumption. We estimated standardized log odds ratios (SLORs) for cases and controls (see Figure S2 in supplementary materials). Each value is the observed log odds ratio for \textit{a pair} of BrS measurements minus the mean LOR from the posterior predictive distribution value, under the model's independence assumption, divided by the standard deviation of the same posterior predictive distribution. We find two large deviations among the cases: RSV with RHINO and RSV with HMPV. These are likely caused by strong seasonality in RSV that is out of phase with weaker seasonality in the other two. Otherwise, the number of SLOR's greater than 2 (8 out of 110) associations is only slightly larger than what is expected under the assumed model (6 expected).

\begin{figure}[!p]
 \centering
\includegraphics[width=\linewidth]{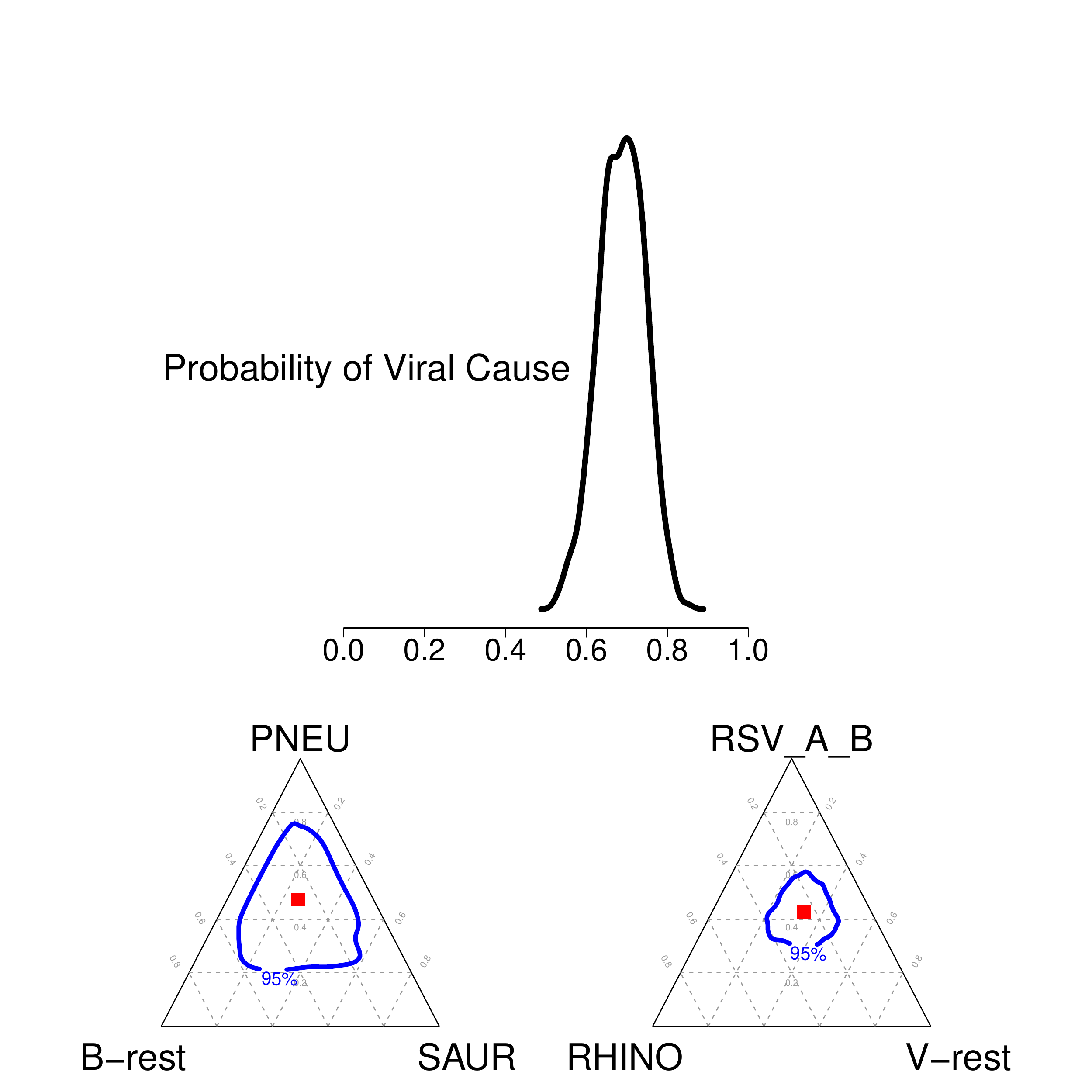}
\caption[Posterior distribution of etiology due to virus or bacteria using expert priors on TPRs.]{Summary of posterior distribution of pneumonia etiology estimates. Top: posterior distribution of viral etiology; bottom left (right): posterior etiology distribution for top two causes given a bacterial (viral) infection. The blue circles are the $95\%$ credible regions \textit{within} the bacterial or viral groups.}
\label{fig:category_pie_WF}
\end{figure}

An attractive feature of using MCMC to estimate posterior distributions is the ease of estimating posteriors for functions of the latent variables and/or parameters. One interesting question from a clinical perspective is whether viruses or bacteria are the major cause and among each subgroup, which species predominate. Figure \ref{fig:category_pie_WF} shows the posterior distribution for the rate of viral pneumonia on the top, and then the conditional distributions of the two leading viruses (bacteria) among viral (bacterial) causes below on the right (left). The posterior distribution of the viral etiologic fraction has mode around $70.0\%$ with $95\%$ credible interval $(57.0\%,79.2\%)$. As shown at the bottom left in Figure \ref{fig:category_pie_WF}, PNEU accounts for most bacterial cases ($47.2\%$ $(24.9\%,71.1\%)$), and SAUR accounts for $25.5\%$ $(8.7\%,49.9\%)$. Of all viral cases (bottom right), RSV is estimated to cause about $42.9\%$ $(32.8\%,54.8\%)$, and RHINO about $24.2\%$ $(13.7\%,37.2\%)$.

\section{Discussion}
\label{sec:discussion}
In this paper, we estimated the frequency with which pathogens cause disease in a case population using a partially-latent class model (pLCM) to allow for known states for a subset of subjects and for multiple types of measurements with different error rates. In a case-control study of disease etiology, measurement error will bias estimates from traditional logistic regression and attributable fraction methods. The pLCM avoids this pitfall and more naturally incorporates multiple sources of data. Here we formulated the model with three levels of measurement error rates.

Absent GS data, we show that the pLCM is only partially identified because of the relationship between the estimated TPR and prevalence of the associated pathogen in the population. Therefore, the inferences are sensitive to the assumptions about the TPR. Uncertainty about their values persists in the final inferences from the pLCM regardless of the number of subjects studied. 

The current model provides a novel solution to the analytic problems raised by the PERCH Study. This paper introduces and applies pLCM to a preliminary set of data from one PERCH study site. Confirmatory laboratory testing, incorporation of additional pathogens, and adjustment for potential confounders may change the scientific findings that will be reported the final complete analysis of the study results when it is completed.

An essential assumption relied upon in the pLCM is that the probability of detecting one pathogen at a peripheral body site depends on whether that pathogen is infecting the child's lung, but is unaffected by the presence of other pathogens in the lung, that is, the non-differential misclassification error assumption. We have formulated the model to include GS measures even though they are available only for a small and unrepresentative subset of the PERCH cases. In general, the availability of GS measures makes it possible to test this assumption as has been discussed by \cite{albert2008estimating}.


Several extensions have potential to improve the quality of inferences drawn and are being developed for PERCH. First, because the control subjects have known class, we can model the dependence structure among the BrS measurements and use this to avoid aspects of the conditional independence assumption central to most LCM methods. The approach is to extend the pLCM to have $K$ subclasses within each of the current disease classes. These subclasses can introduce correlation among the BrS measurements given the true disease state. An interesting question is about the bias-variance trade-off for different values of $K$. This ideas follows previous work on the PARAFAC decomposition of probability distribution for multivariate categorical data \citep{dunson2009nonparametric}. This extension will enable model-based checking of the standard pLCM. 

Second, in our analyses to date, we have assumed that the pneumonia case definition is error-free. Given new biomarkers and availability of chest radiographs that can improve upon the clinical diagnosis of pneumonia, one can introduce an additional latent variable to indicate true disease status and use these measurements to probabilistically assign each subject as a case or control. Finally, regression extensions of the pLCM would allow PERCH investigators to study how the etiology distributions vary with HIV status, age group, and season.

\section*{Acknowledgments}
We thank the members of the larger PERCH Study Group for discussions that helped shape the statistical approach presented herein, and the study participants. We also thank the members of PERCH Expert Group who provided external advice. 


\bibliographystyle{apalike}
\bibliography{refs}

\newpage
\appendix
\section{Full conditional distributions in Gibbs sampler}
\label{appendix:fullconditionals}
In this section, we provide analytic forms of full conditional distributions that are essential for Gibbs sampling algorithm. We use data augmentation scheme by introducing latent lung state $I^L_i$ into the sampling chain and we have the following full conditional distributions: 
\begin{itemize}
\item $\left[I^L_i\mid \text{others}\right]$. If $M^{\GS}_i$ is available, $\text{Pr}\left(I_i^L=j \mid \text{others}\right)= 1$, if $M^{\GS}_{ij}=1$ and $M^{\GS}_{il}=0$, for $l\neq j$; otherwise zero. If $M^{\GS}_i$ is missing, according as whether $M^{\SSs}_i$ is available, the full conditional is given as
\begin{eqnarray}
\text{Pr}(I^L_i=j\mid \text{others})& \propto &\left(\theta_j^{\BrS}\right)^{M^{\BrS}_{ij}}\left(1-\theta^{\BrS}_j\right)^{1-M^{\BrS}_{ij}}\prod_{l\neq j}\left(\psi_l^{\BrS}\right)^{M^{\BrS}_{il}}\left(1-\psi_l^{\BrS}\right)^{1-M^{\BrS}_{il}}\nonumber\\
&&\cdot  \left[\left(\theta_j^{\SSs}\right)^{M^{\SSs}_{ij}}(1-\theta^{\SSs}_j)^{1-M^{\SSs}_{ij}}\mathbf{1}_{\left\{\sum_{l\neq j}M^{\SSs}_{il}=0\right\}}\right]^{\mathbf{1}_{\{j\leq J'\}}}\cdot \pi_j;
\end{eqnarray}
if SS measurement is not available for case $i$, we remove terms involving $M^{\SSs}_{ij}$.
\item $\left[\psi_j^{\BrS}\mid\text{others}\right]\sim \text{Beta}\left(N_j+b_{1j},
n_1-\sum_{i:Y_i=1}\mathbf{1}_{\{I^L_i=j\}}+n_0-N_j+b_{2j}\right)$, where $n_1$ and $n_0$ are number of cases and controls, respectively, and
$N_j = \sum_{i:Y_i=1, I^L_i\neq j}M_{ij}^{\BrS}+\sum_{i:Y_i=0}M^{\BrS}_{ij}$ is the number of positives at position $j$ for cases with $I^L_i\neq j$ and all controls.

\item $\left[\theta_j^{\BrS}\mid\text{others}\right]\sim \text{Beta}\left( S_j+c_{1j}, \sum_{i:Y_i=1} \mathbf{1}_{\{I^L_i=j\}} - S_j+c_{2j}\right)$, where $S_j=\sum_{i:Y_i=1,I^L_i=j}M^{\BrS}_{ij}$
is the number of positives for cases with $j$th pathogen as their causes.

\item $\left[\theta_j^{\SSs}\mid\text{others}\right]\sim \text{Beta}\left(T_j+d_{1j},\sum_{i:Y_i=1, \SSs\text{available}}\mathbf{1}_{\{I^L_i=j\}}-T_j+d_{2j}\right)$,
where $T_j = \sum_{i:Y_i=1, I^L_i=j, \SSs\text{available}}M^{\SSs}_{ij}.$ When no SS data is available, this conditional distribution reduces to $\text{Beta}(d_{1j},d_{2j})$, the prior.

\item $\left[\bm{\pi}\mid I_{i}^L,i:Y_i=1\right] \sim
\text{Dirichlet}(a_1+U_1,...,a_J+U_J)$,
where $U_j=\sum_{i:Y_i=1}\mathbf{1}_{\{I^L_i=j\}}$.
\end{itemize}

\section{Pathogen names and their abbreviations}
\label{appendix:pathname}
\textbf{Bacteria}: HINF- \textit{Haemophilus influenzae}; PNEU-\textit{Streptococcus pneumoniae}; SASP-\textit{Salmonella} species; SAUR-\textit{Staphylococcus aureus}. \textbf{Viruses}: ADENO-adenovirus; COR$\_$43-coronavirus OC43; FLU$\_$C-influenza virus type C; HMPV$\_$A$\_$B-human metapneumovirus type A or B; PARA1-parainfluenza type 1 virus; RHINO-rhonovirus; RSV$\_$A$\_$B-respiratory syncytial virus type A or B.

\begin{figure}[!p]
\begin{center}
\includegraphics[width=\textwidth]{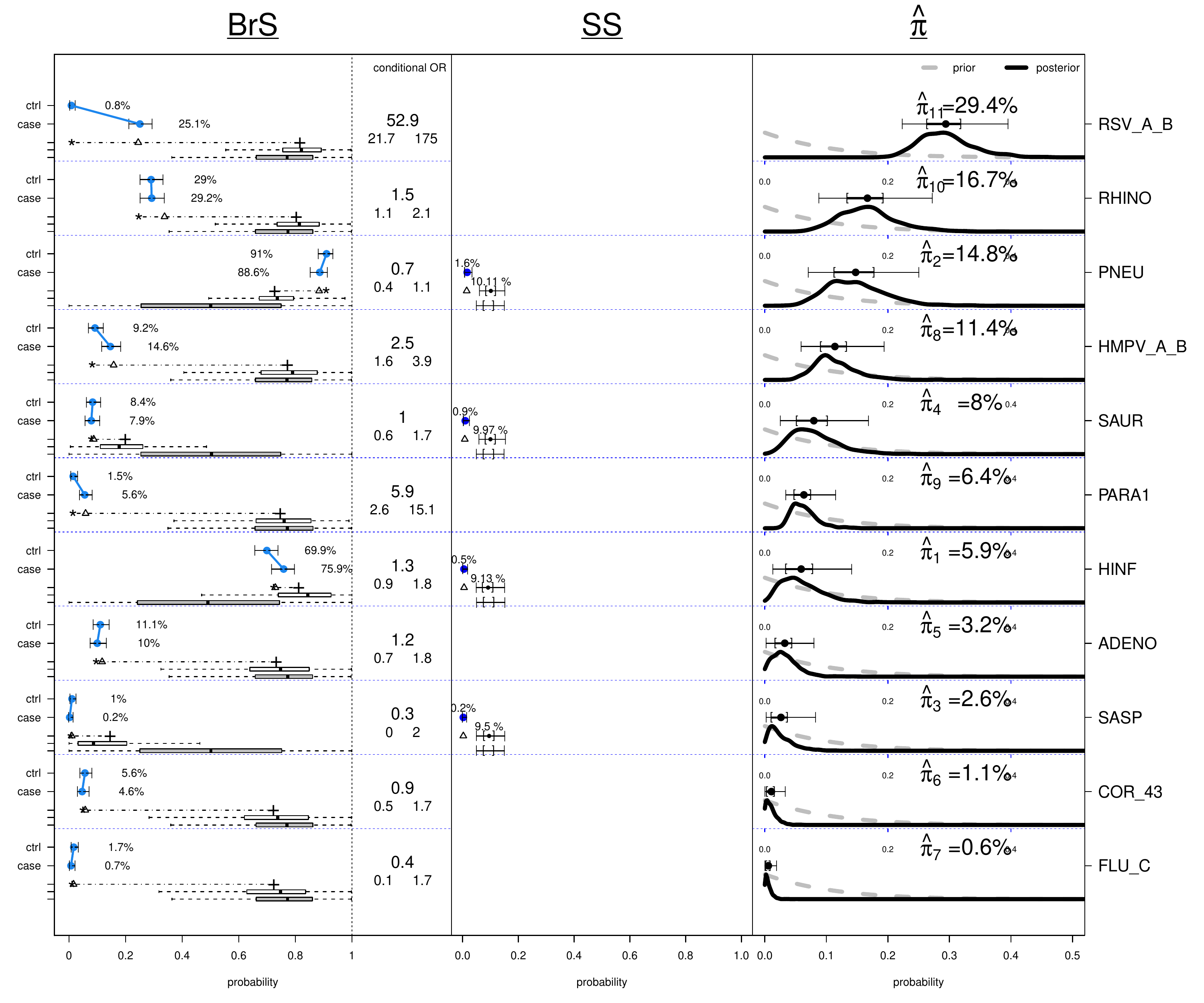}
\end{center}
\vspace{-0.1in}
\caption[Results using expert priors on TPRs.]{The observed BrS rates (with $95\%$ confidence intervals) for cases and controls are shown on the far left. The conditional odds ratio given the other pathogens is listed with $95\%$ confidence interval in the box to the right of the BrS data summary. In the left box, below the case and control observed rates is a horizontal line with a triangle. The line starts on the left at the model estimated false positive rate (FPR, $\hat{\psi}_j^{\BrS}$) and ends on the right at the estimated true positive rate (TPR, $\hat{\theta}_j^{\BrS}$). Below the TPR are two boxplots summarizing its posterior (top) and prior (bottom) distributions. The location of the triangle, expressed as a fraction of the distance from the estimated FPR to the TPR, is the point estimate of the etiologic fraction for each pathogen. The SS data are shown in a similar fashion to the right of the BrS data using support intervals rather than boxplots.}
\label{fig:site02GAM}
\end{figure}

\section*{
Web-based supplementary materials for ``Partially-Latent Class Models (pLCM) for Case-Control Studies of Childhood Pneumonia Etiology"}
Z.Wu {\it et al.}

 \renewcommand{\thefigure}{S\arabic{figure}}
 \setcounter{figure}{0}

\begin{figure}[!htp]
\begin{center}
\includegraphics[width=.8\textwidth]{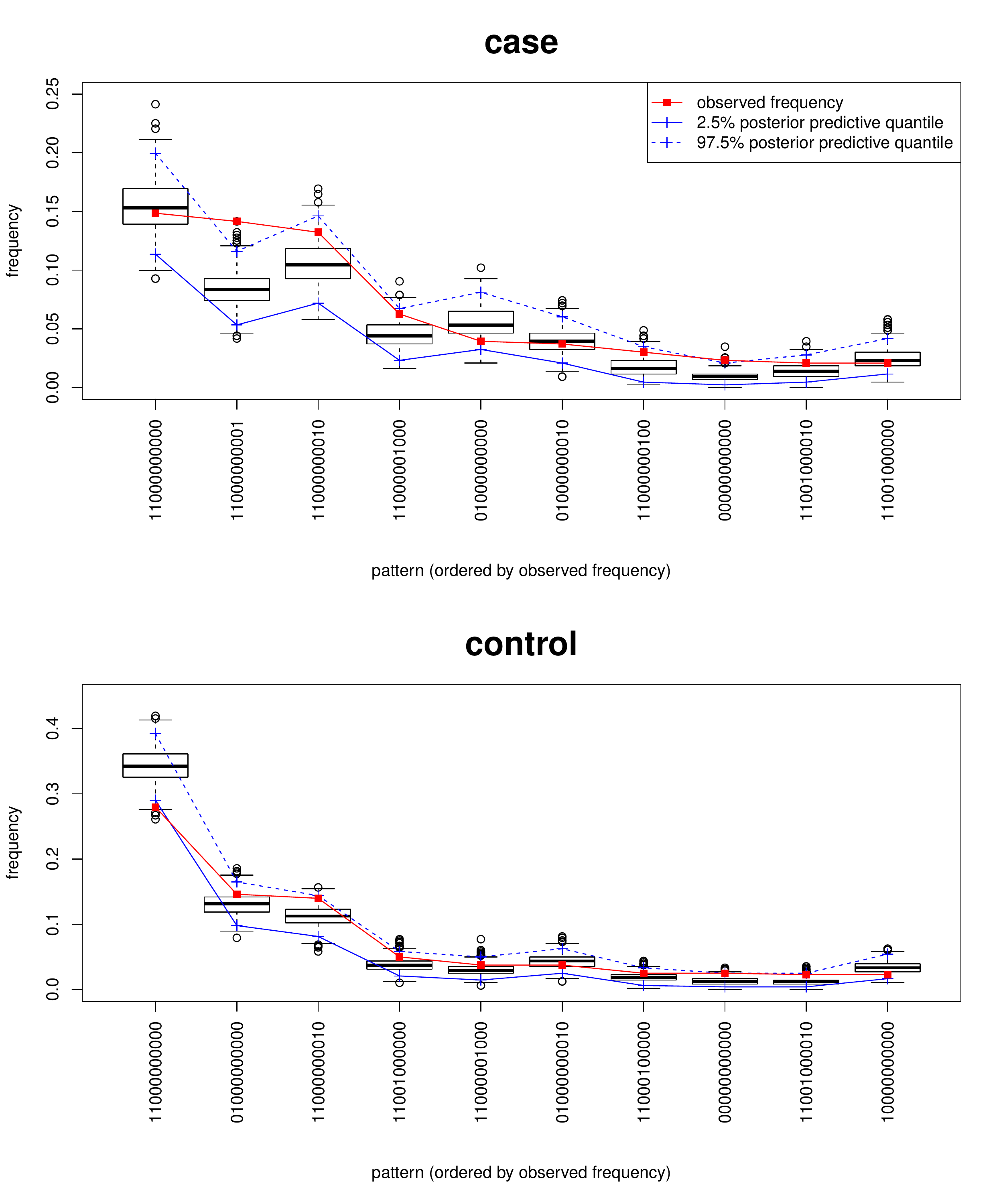}
\end{center}
\vspace{-0.1in}
\caption[Model checking using most frequent measurement patterns]{Posterior predictive checking for 10 most frequent BrS measurement patterns among cases and controls with expert priors on TPRs.}
\label{fig:freq.check}
\end{figure}

\begin{figure}[!p]
\begin{center}
\includegraphics[width=.9\textwidth]{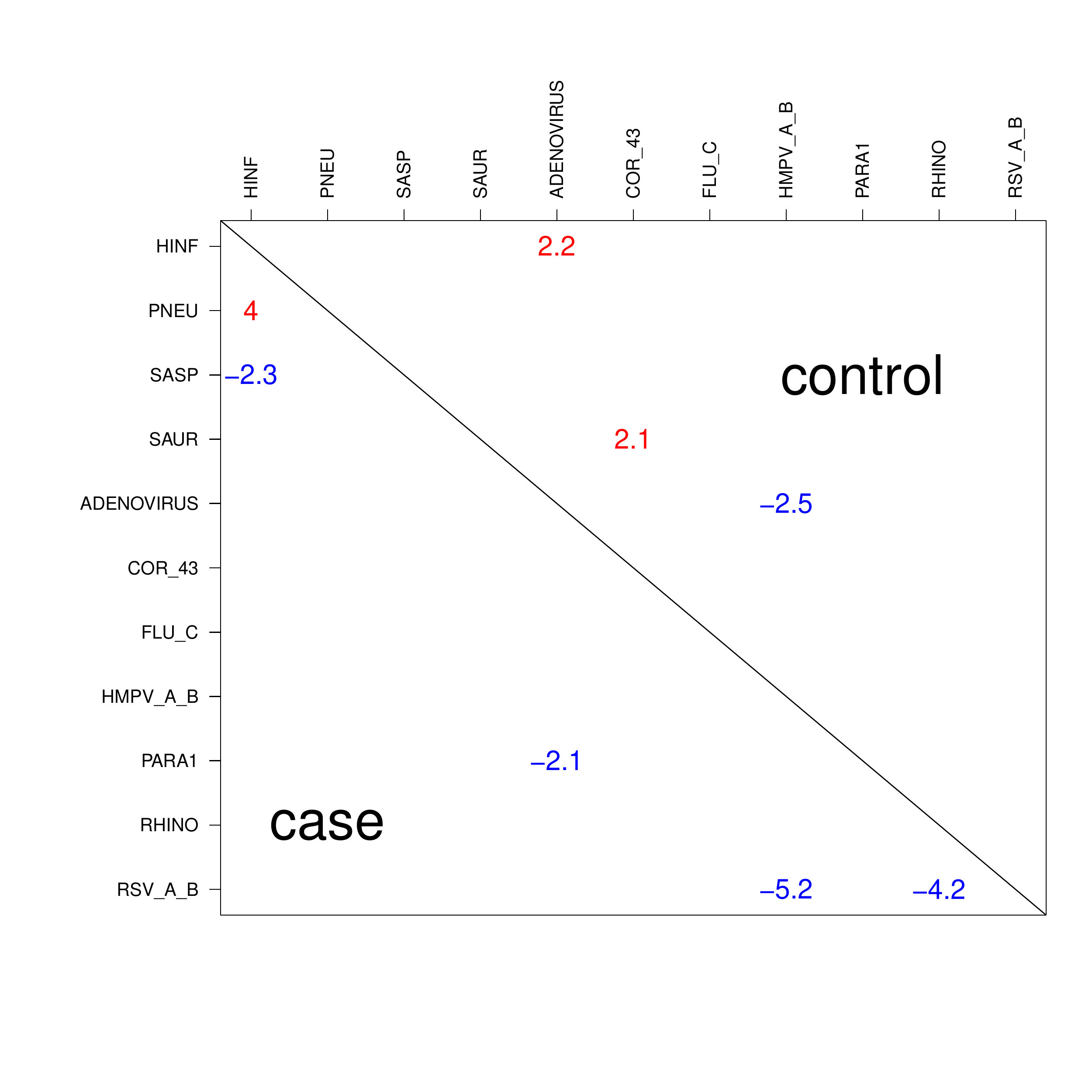}
\end{center}
\vspace{-0.1in}
\caption[Model checking for pairwise odds ratios]{Posterior predictive checking for pairwise odds ratios separately for cases (lower triangle) and controls (upper triangle) with expert priors on TPRs. Each entry is a standardized log odds ratio (SLOR): the observed log odds ratio for a pair of BrS measurements minus the mean LOR for the posterior predictive distribution divided by the standard deviation of the posterior predictive distribution. The first significant digit of absolute SLORs are shown in red for positive and blue for negative values, and only those greater than 2 are shown.}
\label{fig:or.check}
\end{figure}

\end{document}